%% file: apssamp.tex
\documentclass[%
 reprint,
superscriptaddress,
 amsmath,amssymb,
aps,
]{revtex4-2}

\usepackage{graphicx}
\usepackage{dcolumn}
\usepackage{bm}
\usepackage{multirow}
\usepackage{makecell}
\usepackage{float}
\usepackage{xcolor}

\usepackage{units}
\usepackage{comment}
\usepackage{color}
\usepackage{url}

\usepackage[colorlinks]{hyperref}
\hypersetup{%
	plainpages=true,
	breaklinks=true,
	hypertexnames=false,
	pageanchor=true,
	colorlinks=true,
	linkcolor={blue},
	citecolor={red},
	urlcolor={blue},
	anchorcolor={black}
}

\usepackage{mleftright} 

\newcommand{\figref}[1]{\mbox{Fig.~\ref{#1}}}
\newcommand{\figpanel}[2]{Fig.~\hyperref[#1]{\ref*{#1}(#2)}}
\newcommand{\figpanels}[3]{Fig.~\hyperref[#1]{\ref*{#1}(#2)-(#3)}}
\newcommand{\figpanelNoPrefix}[2]{\hyperref[#1]{\ref*{#1}(#2)}}

\renewcommand{\eqref}[1]{\mbox{Eq.~(\ref{#1})}}
\newcommand{\tabref}[1]{\mbox{Table~\ref{#1}}}

\newcommand{\bra}[1]{\mleft\langle #1 \mright |}
\newcommand{\ket}[1]{\mleft|#1 \mright \rangle}

\newcommand{\ketbra}[2]{\mleft| #1 \rangle \langle #2 \mright|}

\newcommand{\be}{\begin{equation}}
\newcommand{\ee}{\end{equation}}
\newcommand{\bea}{\begin{eqnarray}}
\newcommand{\eea}{\end{eqnarray}}

\usepackage{soul}

\usepackage{lineno}

\begin{document}


\preprint{APS/123-QED}

\title{Experimental realization of deterministic and selective photon addition in a bosonic mode assisted by an ancillary qubit}

\author{Marina Kudra}
\affiliation{Department of Microtechnology and Nanoscience, Chalmers University of Technology, 412 96 Gothenburg, Sweden}

\author{Martin Jirlow}
\affiliation{Department of Microtechnology and Nanoscience, Chalmers University of Technology, 412 96 Gothenburg, Sweden}

\author{Mikael Kervinen}
\affiliation{Department of Microtechnology and Nanoscience, Chalmers University of Technology, 412 96 Gothenburg, Sweden}

\author{Axel M. Eriksson}
\affiliation{Department of Microtechnology and Nanoscience, Chalmers University of Technology, 412 96 Gothenburg, Sweden}

\author{Fernando Quijandr\'{\i}a}
\affiliation{Quantum Machines Unit, Okinawa Institute of Science and Technology Graduate University, Onna-son, Okinawa 904-0495, Japan}

\author{Per Delsing}
\affiliation{Department of Microtechnology and Nanoscience, Chalmers University of Technology, 412 96 Gothenburg, Sweden}

\author{Tahereh Abad}
\email{tahereh.abad@chalmers.se}
\affiliation{Department of Microtechnology and Nanoscience, Chalmers University of Technology, 412 96 Gothenburg, Sweden}

\author{Simone Gasparinetti}
\email{simoneg@chalmers.se}
\affiliation{Department of Microtechnology and Nanoscience, Chalmers University of Technology, 412 96 Gothenburg, Sweden}

\date{\today}

\begin{abstract}

Bosonic quantum error correcting codes are primarily designed to protect against single-photon loss. To correct for this type of error, one can encode the logical qubit in code spaces with a definite photon parity, such as cat codes or binomial codes. Error correction requires a recovery operation that maps the error states -- which have opposite parity -- back onto the code space. Here, we realize a collection of photon-number-selective, simultaneous photon addition operations on a bosonic mode, a microwave cavity, assisted by a superconducting qubit. These operations are implemented as two-photon transitions that excite the cavity and the qubit at the same time. The additional degree of freedom of the qubit makes it possible to implement a coherent, unidirectional mapping between spaces of opposite photon parity. We present the successful experimental implementation of the drives and the phase control they enable on superpositions of Fock states. The presented technique, when supplemented with qubit reset, is suitable for autonomous quantum error correction in bosonic systems, and, more generally, opens the possibility to realize a range of non-unitary transformations on a bosonic mode.

\end{abstract}


\maketitle

\section{\label{sec:Introduction}Introduction}
Encoding quantum information in bosonic modes, such as harmonic oscillators, is a promising platform for error correctable quantum computing \cite{joshi2021quantum}. However, in order to control the quantum information in a linear mode beyond Gaussian operations some nonlinearity is required, which is commonly realized by an ancilla qubit dispersively coupled to the bosonic mode. The dispersive interaction enables an indirect control of the oscillator states and allows for different gates to be implemented. Examples are the Selective Number-dependent Arbitrary Phase (SNAP) gate~\cite{heeres2015cavity,krastanov2015universal,PRXQuantum.3.030301} that changes only the phases of the Fock states, Echoed Conditional Displacement (ECD)~\cite{eickbusch2022fast} that generates conditional displacements in phase-space, and optimal control pulses~\cite{heeres2017implementing} that in general can change both the phases and the populations of the bosonic mode.

Precise control of the bosonic mode enables quantum error correction against unwanted or uncontrolled errors \cite{cai2021bosonic}. Often, the dominant source of error in a bosonic mode is single-photon loss. Bosonic codes such as the cat and binomial codes remain invariant under discrete-order rotations in phase space. The order of such a rotation determines their Fock space structure and consequently their robustness against photon loss and gain errors. For instance, the four-component cat code~\cite{Mirrahimi2014}, i.e., a superposition of four orthogonal coherent states, is the smallest-order version of the cat code which is robust against single-photon losses~\cite{Grimsmo2020RSB}.
The logical words of the cat code are parity eigenstates and therefore by periodically monitoring the parity of the system we find whether an error has happened or not.
Ofek et.al.~\cite{ofek2016extending} successfully implemented the real-time parity detection of cat codes and subsequent active reset of the qubit, but they did not correct the deterministic loss of amplitude of the cat states. 
Another approach is to use the binomial code~\cite{ni2022beating,hu2019quantum} and after parity detection use optimal control pulses synthesized by GRAPE~\cite{heeres2017implementing} to go back to the code space. 
An alternative to the above protocols corresponds to the recently introduced Parity Recovery by Selective Photon Addition (PReSPA)~\cite{gertler2021protecting}. This is an autonomous, measurement-free error correction scheme realized by means of number-selective continuous wave driving fields. PReSPA is implemented by driving two sets of microwave combs, both of which have to be photon-number selective. This is a disadvantage as it restricts the rates of all the drives. If we were to implement PReSPA in a pulsed version, both pulses would have to be long compared to the inverse of the dispersive shift. We propose to implement the same autonomous recovery of parity by applying a Selective Number-dependent Arbitrary-Phase Photon-Addition (SNAPPA) gate followed by a fast unconditional qubit reset.

The SNAPPA gate is described by
\begin{equation} \label{SNAPPA_gate}
S_{\rm PA}(\{(\theta_i)_{n_i \to n_{i+1}}\}) \equiv \sum_{i\subset N} e^{i\theta_{i}} \ket{n_i+1}\ket{e} \bra{n_i}\bra{g} + {\rm h.c.},
\end{equation}
where $\ket{n_i}$ are the Fock states which are affected by the transformation when the qubit is in the ground state, $\{\theta_i\}$ the corresponding phases they acquire,
and $\ket{g}$ and $\ket{e}$ are the ground and excited states of the qubit, respectively.
We implement this gate by driving a two-photon transition of the combined qubit-cavity system. We show that we can selectively add photons to the odd-parity subspace without affecting the even-parity subspace. Moreover, we control the relative phases of the added photons by adjusting the phases of the off-resonant drives. The two-photon transition $\ket{0}\ket{g}\rightarrow \ket{1}\ket{e}$ was previously implemented in order to create entanglement between two distant transmon qubits~\cite{Campagne-Ibarcq2018}. A special case of the SNAPPA gate, where the transitions for all Fock states $\ket{n}$ are driven simultaneously and $\theta_n=0$ for all $n$ corresponds to the so-called 'conventional' single-photon addition  on the bosonic mode, which was implemented in trapped ions~\cite{Um2016}. The latter operation is not to be confused with the bosonic ladder operator $\hat{a}^\dag$ \cite{Parigi2007}. 

\begin{figure}[t]
\includegraphics[width=70mm]{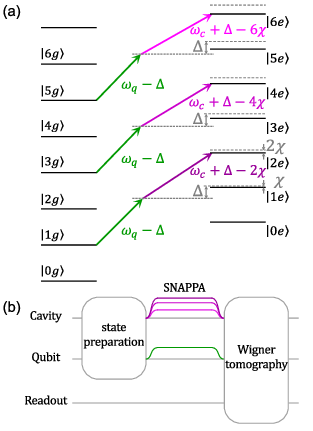}
\caption{\label{fig:setup} (a) Level diagram sketching the SNAPPA gate, here implemented to map the cavity odd parity subspace onto the even-parity subspace, 
without compromising the even-parity states.
The qubit off-resonant drive at frequency $\omega_q-\Delta$ is common for all the transitions(green arrows). In different shades of magenta arrows, the cavity off-resonant drives at frequencies $\omega_c+\Delta-(n+1)\chi$ enable the transition from $\ket{n}\ket{g}$ to $e^{i\theta_n}\ket{n+1}\ket{e}$, for each odd $n$ respectively. The phase $\theta_n$ is directly related to the relative phases of the drives $\omega_c+\Delta-(n+1)\chi$ (b) Pulse sequence used to apply the SNAPPA gate. We initialize the cavity and the qubit to $\ket{\Psi}\ket{g}$, apply the SNAPPA gate, and perform direct Wigner tomography to measure the state of the cavity.
}
\end{figure}

\section{Methods}

\begin{figure*}[t!]
\includegraphics[width=140mm]{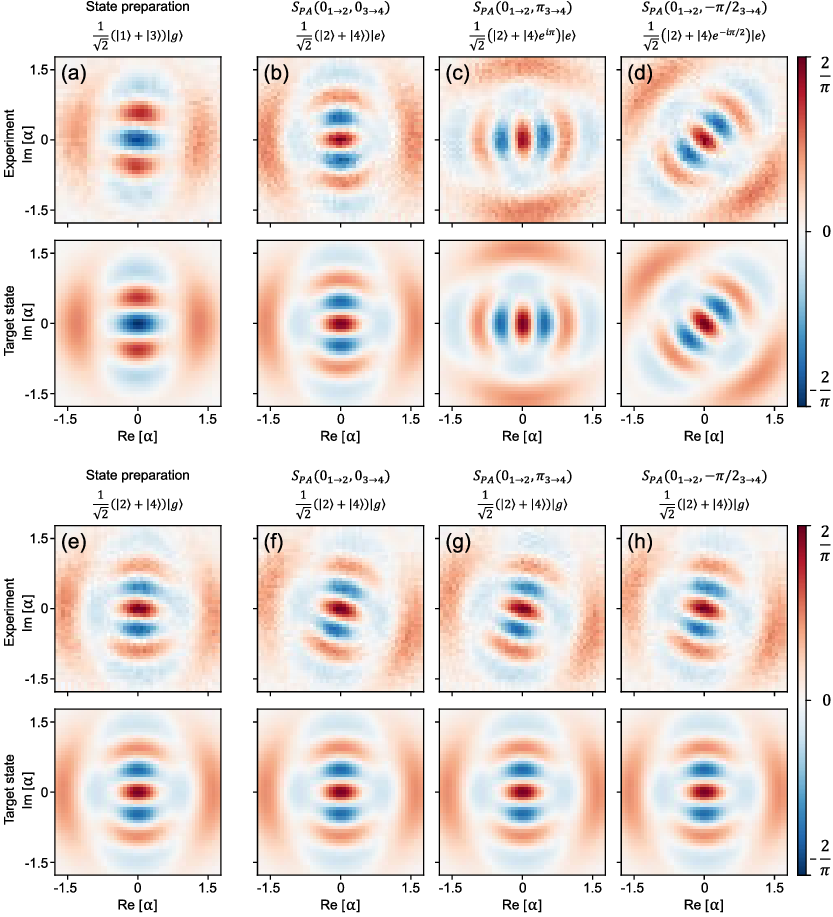}
\caption{\label{fig:wig_13} Characterization of the SNAPPA gate $S_{\rm PA}(0_{1\rightarrow2},\theta_{3\rightarrow4})$ to map odd-parity states, i.e., $\{\ket{1}, \ket{3}\}$ to even-parity states, i.e., $\{\ket{2}, \ket{4}\}$, while the even states are almost unaffected by the SNAPPA gate. Wigner functions of experimental versus target states when the SNAPPA gate is applied to (a-d) an error initial state
and (e-h) a computational initial state.
 (a) Initial state $(\ket{1}+\ket{3})\ket{g}/\sqrt{2}$ and final/target states $(\ket{2}+\ket{4}e^{i\theta_{3\rightarrow4}})\ket{e}/\sqrt{2}$, where $\theta_{3\rightarrow4}$ is the phase of the cavity off-resonant drive $\omega_c+\Delta-4\chi$, given by (b)~$\theta_{3\rightarrow4}=0$, (c)~$\theta_{3\rightarrow4}=\pi$, (d)~$\theta_{3\rightarrow4}=\pi/2$. The lower panels show that the drives have little to no effect on the even-parity subspace. (e) Initial state $(\ket{2}+\ket{4})\ket{g}/\sqrt{2}$ and (f)-(h) identical final state $(\ket{2}+\ket{4})\ket{g}/\sqrt{2}$ for $\theta_{3\rightarrow4} = 0,\, \pi,-\pi/2$.}
\end{figure*}


We implement SNAPPA in a circuit quantum electrodynamics setup comprising a long-lived microwave cavity \cite{reagor2016,kudra2020high}, a dispersively coupled transmon qubit, and a readout resonator. We cool down the device, which is similar to Ref.~\cite{PRXQuantum.3.030301}, in a dilution refrigerator with base temperature below 10~mK. We control the whole system with a single microwave transceiver that uses direct digital synthesis to generate microwave pulses up to 8.5~GHz with an instantaneous bandwith of 1~GHz~\cite{Mats2022}.

The relevant part of the qubit-cavity system is described by the Hamiltonian~\cite{heeres2015cavity}
\begin{eqnarray}
\label{eq:hamil}
    H &=& \omega_c a^\dagger a - \dfrac{K_c}{2}a^{\dagger2} a^2 + \omega_{q} q^\dagger q- \dfrac{\alpha_q}{2}q^{\dagger2} q^2 \nonumber \\
    &-& \chi a^\dagger a \, q^\dagger q - \dfrac{\chi'}{2}a^{\dagger2} a^2 q^\dagger q,
\end{eqnarray}
where $\omega_c$ and $\omega_q$ are the resonance frequencies of the cavity and the transmon qubit, $K_c$ is the Kerr nonlinearity of the cavity, $\alpha_q$ is the qubit anharmonicity, $\chi$ is the dispersive shift between the cavity and the qubit, $\chi'$ is a photon-number-dependent correction to the dispersive shift, $a^\dagger$ ($a$) is the creation (annihilation) operator of the cavity field and similarly, and $q^\dagger$ ($q$) is the raising (lowering) operator of the qubit. The Hamiltonian parameters and the coherence properties of the system are listed in~\cite{SuppMat}. 

\begin{figure*}[t!]
\includegraphics[width=180mm]{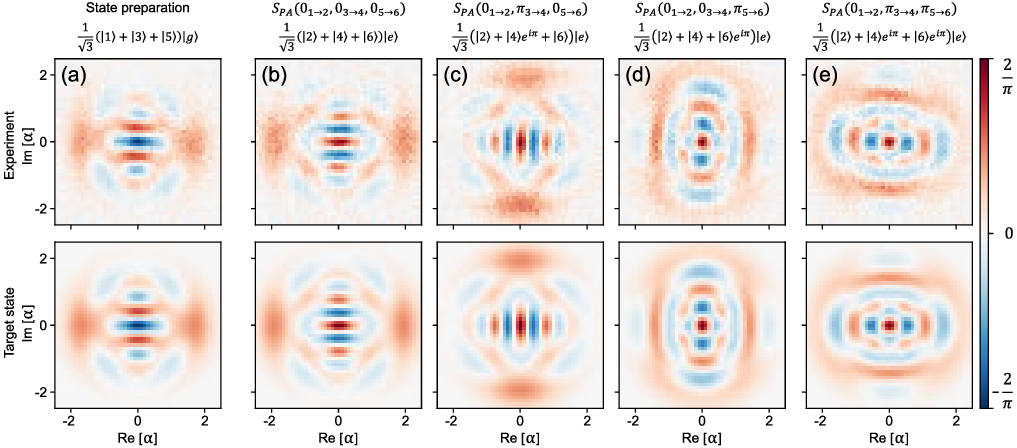}
\caption{\label{fig:wig_135}Wigner functions of experimentally obtained and target states when SNAPPA gate $S_{\rm PA}(0_{1\rightarrow2},\theta_{3\rightarrow4},\theta_{5\rightarrow6})$ is applied. (a) Initial state  $1/\sqrt{3}(\ket{1}+\ket{3}+\ket{5})\ket{g}$ and the resulting states after one iteration of applying the drives $(\ket{2}+\ket{4}e^{i\theta_{3\rightarrow4}}+\ket{6}e^{i\theta_{5\rightarrow6}})\ket{e}/\sqrt{3}$, where $\left(\theta_{3\rightarrow4}, \theta_{5\rightarrow6} \right)$ is the phase of cavity off-resonant drive with frequency $\omega_c+\Delta-4\chi$ and $\omega_c+\Delta-6\chi$, respectively, given by (b)~$(0,0)$, (c)~$(\pi,0)$, (d)~$(0,\pi)$ and (e)~$(\pi,\pi)$.}
\end{figure*}


We implement SNAPPA by applying off-resonant drives to the qubit and the cavity, whose frequencies sum up to implement the two-excitation transitions $\ket{n}\ket{g}$ to $\ket{n+1}\ket{e}$ at frequencies $\omega_c+\omega_q-(n+1)\chi$. We apply one qubit drive at frequency $\omega_q-\Delta$, where $\Delta$ is detuning between the qubit transition frequency and the drive frequency and a frequency comb to the cavity at frequencies $\{\omega_c+\Delta-(n_i+1)\chi\}$, with one component for each transition $\ket{n_i}\ket{g}$ to $\ket{n_i+1}\ket{e}$. Which states $\{n_i\}$ are addressed is chosen by applying the appropriate frequency of the cavity frequency comb. The phase $\theta_i$ that will be added to the final Fock state $\ket{n_i+1}$ is determined by the relative phases of the drive at the frequency $\omega_c+\Delta-(n_i+1)\chi$. The level diagram in \figref{fig:setup} illustrates the SNAPPA gate
$S_{\rm PA}(\theta_{1\rightarrow2},\theta_{3\rightarrow4},\theta_{5\rightarrow6})$,
which requires one frequency tone to the qubit (green arrows) and three tones to the cavity (in shades of magenta arrows).

When choosing $\Delta$, the condition $|\Delta-(n_{\rm max}+1)\chi|\gg 1/\tau$ (where $\tau$ is the gate time) has to hold, so that the pulses we send are not driving cavity displacements or qubit rotations directly. In our experiments we simultaneously address up to 3 transition with a maximum Fock number of 5.
On the other hand, $\Delta$ should be smaller than qubit anharmonicity $|\alpha_b|$ to avoid driving the first-to-second-excited-state qubit transition. A smaller $\Delta$ is also preferable to achieve the same effective strength with less power in the driving tones.
Theoretically, driving the qubit at the frequency $\omega_q + \Delta$ should avoid this restricition. Nevertheless, our experimental attempts did not yield the desired results. Based on an analysis of the effective Hamiltonian~\cite{Jirlow2025}, we suspect the presence of a destructive interference effect arising from Stark shifts.
The pulses have to be selective depending on the Fock number in the cavity, thus we require $\tau \gg 1/\chi$. At the same time, $\tau$ should be chosen as short as possible to avoid qubit dephasing and decoherence during the operation.
Based on these tradeoffs, we choose pulses with length of $\tau=$ \unit[4.2]{$\mu$s} with a \unit[100]{ns} $\sin^2$ rise- and fall-time for $n_{\rm max}=5$, $\chi=$ \unit[1.4]{MHz}, and $\Delta/(2\pi)=$ \unit[30]{MHz}.
 
The off-resonant drives induce Stark shifts of both the qubit and the cavity frequencies and hence the two-excitation transitions require a frequency correction $d\omega_n$, which we include in the cavity off-resonant drives $\omega_c+\Delta-(n+1)\chi+d\omega_n$, where $d\omega_n/(2\pi)$ is of the order of a few hundred kHz and is calibrated together with the amplitudes and phases of all the drives in the self-consistent procedure we explain in \cite{SuppMat}.

To characterize SNAPPA, we initially prepare a range of nonclassical states using a sequence of three displacements and two optimized selective number-dependent arbitrary phase (SNAP) gates~\cite{PRXQuantum.3.030301}. 
After the gate, we perform direct Wigner tomography of the cavity mode~\cite{heeres2015cavity}, where we apply a displacement of varying complex amplitude $\alpha$ followed by a Ramsey measurement at a fixed time delay of $1/(2\chi) \approx$~\unit[360]{ns}, which returns the photon parity of the cavity mode as the outcome of the qubit measurement. The Wigner function is obtained as
\begin{equation}
    W(\alpha) = \frac{2}{\pi}\mathrm{Tr}[D^\dagger(\alpha) \rho D(\alpha) \Pi],
\end{equation}
where $\Pi$ is the parity operator.
We reconstruct the density matrix of the target state from the Wigner function using a recently developed neural-network-based approach~\cite{Ahmed2021a, Ahmed2021b} which reconstructs the density matrix of the state by minimizing a cost function between the reconstructed state and the measured Wigner tomography data.

\begin{table}[h]
\caption{\label{tab:Qubit_populations_and_fidelities}
Qubit population and fidelity to target state. Fidelities marked with * are fidelities to the rotated state $(\ket{2}+\ket{4}e^{i0.53})\ket{g}/\sqrt{2}$.}
\begin{ruledtabular}
\begin{tabular}{cccc}
Fig.&Target & Qubit  & Fidelity\\ 
 & state &   population & bosonic state\\
\hline
2(a)&$1/\sqrt{2}(\ket{1}+\ket{3})\ket{g}$ & 0.04 & 0.96 \\
2(b)&$1/\sqrt{2}(\ket{2}+\ket{4})\ket{e}$ & 0.92 & 0.94 \label{2b}\\
2(c)&$1/\sqrt{2}(\ket{2}+\ket{4}e^{i\pi})\ket{e}$ & - & 0.93 \\
2(d)&$1/\sqrt{2}(\ket{2}+\ket{4}e^{-i\pi/2})\ket{e}$ & - & 0.94 \\

2(e)&$1/\sqrt{2}(\ket{2}+\ket{4})\ket{g}$ & 0.06 & 0.93 \\
2(f)&$1/\sqrt{2}(\ket{2}+\ket{4})\ket{g}$ & 0.09 & 0.92* \\
2(g)&$1/\sqrt{2}(\ket{2}+\ket{4})\ket{g}$ & - & 0.93* \\ 
2(h)&$1/\sqrt{2}(\ket{2}+\ket{4})\ket{g}$ & - & 0.92* \\
\\
3(a)&$1/\sqrt{3}(\ket{1}+\ket{3}+\ket{5})\ket{g}$ & 0.05 & 0.92 \\
3(b)&$1/\sqrt{3}(\ket{2}+\ket{4}+\ket{6})\ket{e}$ & 0.90 & 0.93 \\
3(c)&$1/\sqrt{3}(\ket{2}+\ket{4}e^{i\pi}+\ket{6})\ket{e}$ & - & 0.93 \\
3(d)&$1/\sqrt{3}(\ket{2}+\ket{4}+\ket{6}e^{i\pi})\ket{e}$ & - & 0.93 \\
3(e)&$1/\sqrt{3}(\ket{2}+\ket{4}e^{i\pi}+\ket{6}e^{i\pi})\ket{e}$ & - & 0.93 \\
\\
4(a)&$1/\sqrt{2}(\ket{0}+\ket{1}e^{i\pi})\ket{g}$ & 0.02 & 0.93 \\
4(b)&$1/\sqrt{2}(\ket{1}+\ket{2}e^{i\pi})\ket{e}$ & 0.94 & 0.92 \\
4(c)&$(\sqrt{2}/\sqrt{3}\ket{1}+1/\sqrt{3}\ket{3})\ket{g}$ & 0.05 & 0.90 \\
4(d)&$(\sqrt{2}/\sqrt{3}\ket{2}+1/\sqrt{4}\ket{3})\ket{e}$ & 0.91 & 0.96 \\
\end{tabular}
\end{ruledtabular}
\end{table}

\section{Results}
\subsection{Using SNAPPA to recover parity with controllable relative phase}

For logical states with a fixed parity, single-photon loss projects the states into orthogonal parity subspace. Thus we use two-excitation selective transitions to project the states back into the logical states. The two-excitation selective transitions have to map the phase of odd Fock states (error) to their even counterparts (computational) and excite the qubit. At the same time, they should keep the even-parity subspace unchanged and leave the qubit in the ground state. 

We evaluate these requirements by preparing $\left(\ket{1}+\ket{3} \right)\ket{g}/\sqrt{2}$ and $\left(\ket{2}+\ket{4} \right)\ket{g}/\sqrt{2}$ shown in \figpanel{fig:wig_13}{a}, and~\figpanel{fig:wig_13}{e}, respectively. We apply the SNAPPA gate $S_{\rm PA}(0_{1\rightarrow2},\theta_{3\rightarrow4})$ for $\theta_{3\rightarrow4} = 0,\, \pi,-\pi/2$ by applying a qubit off-resonant drive at frequency $\omega_q-\Delta$ and two cavity off-resonant drives at frequencies $\omega_c+\Delta-2\chi$ and $\omega_c+\Delta-4\chi$ in order to drive transitions $\ket{1}\ket{g}\rightarrow\ket{2}\ket{e}$ and $\ket{3}\ket{g}\rightarrow\ket{4}\ket{e}$ simultaneously. We then vary the phase of the second cavity drive at frequency $\omega_c+\Delta-4\chi$ to vary the phase $\theta_{3\rightarrow4}$.

Qubit population and fidelity of the cavity state are listed in Table~\ref{tab:Qubit_populations_and_fidelities}.
Because we perform Wigner tomography directly after applying the state preparation gates and the SNAPPA gates, the reported fidelities suffer from different type of errors due to residual qubit population and qubit-cavity entanglement. We measure an average excited state population of the qubit of about 5\% after preparing the initial states. We also observe a reduced contrast in the Ramsey sequence used for Wigner tomography, corresponding to an additional 3 to 4\% change in qubit population, due to the SNAPPA gate. Resetting the qubit prior to the tomography step would lead to a more accurate estimate of the fidelity.
In \cite{SuppMat}, we show that any initial population in the qubit leads to a direct misassignment of the measured photon-number parity, with the error probability equal to the initial qubit excitation. This highlights the importance of qubit reset before tomography to ensure an accurate reconstruction of the Wigner function, which we leave for future work.

We analyze the error budget of the SNAPPA gate by numerically simulating the transitions $\ket{1}\ket{g} \rightarrow \ket{2}\ket{e}$ and $\ket{3}\ket{g} \rightarrow \ket{4}\ket{e}$ simultaneously, both with and without decoherence error channels. Additionally, we simulate the system with incoherent errors only in either the qubit or cavity, respectively, to assess the error contributions from each component. The drive parameters are numerically optimized to maximize the fidelity of the target state. The fidelity of the simulated bosonic state is calculated by tracing out the qubit and determining the overlap with the target state.

The simulated qubit populations and cavity state fidelities are presented in Table \ref{tab:Simulated_qubit_populations_and_fidelities}. The results indicate that qubit decoherence is the dominant incoherent error. The dive parameters used in the simulation were optimized assuming the absence of losses, which explains the larger infidelities observed here compared to the experimental results presented in Table \ref{tab:Qubit_populations_and_fidelities}.

Note that while coherent errors due to entanglement also exist, Table \ref{tab:Simulated_qubit_populations_and_fidelities} indicates that the qubit relaxation is the dominant contribution.

\begin{table}[h]
    \centering
    
    \caption{Simulated error budget from simulating the $\ket{1}\ket{g} \rightarrow \ket{2}\ket{e}$ and $\ket{3}\ket{g} \rightarrow \ket{4}\ket{e}$ transitions simulateneously, with and without decoherence.}
    \begin{ruledtabular}
    \begin{tabular}{c|cccc}
        Included error & Qubit & Fidelity \\
        & population & bosonic state \\
        \hline
        No error & 0.98 & 0.97 \\
        Qubit relaxation & 0.88 & 0.87 \\
        Cavity relaxation & 0.96 & 0.94 \\
        Both qubit \& cavity & 0.87 & 0.85 \\
    \end{tabular}
    \end{ruledtabular}
    
    \label{tab:Simulated_qubit_populations_and_fidelities}
\end{table}


Abrupt transition the phase of the second cavity drive, we directly change the phase of Fock state $\ket{4}$ for the odd-parity initial state (\figpanels{fig:wig_13}{b}{d}). The fidelity is 0.02-0.03 lower than the state preparation fidelity. The observed reduction in fidelity following the application of SNAPPA can likely be attributed to two factors. First, during SNAPPA drives, the injection of photons into the cavity will increase its susceptibility to loss errors. Second, fidelity diminishes during Wigner tomography when the qubit starts in its excited state, as it is more susceptible to decoherence.

As seen in \tabref{tab:Qubit_populations_and_fidelities}, particularly after applying the SNAPPA gate, the qubit error is larger than the cavity error for transitions that excites the qubit due to the stronger decay rate of the qubit. As mentioned above, the dominant source of error in SNAPPA is qubit relaxation. During the gate operation, transitions such as $\left(\ket{1}+\ket{3} \right)\ket{g}/\sqrt{2}$ to $\left(\ket{2}+\ket{4} \right)\ket{g}/\sqrt{2}$, shown in Table \ref{tab:Qubit_populations_and_fidelities} 2(b), are driven by tuning the drive parameters to induce Rabi oscillations. At the point of full transition, which we set to occur at 4.2 $\mu$s, a single photon is already being injected into the cavity, and the qubit is expected to be in its excited state. If the qubit decays during this period, the resulting incoherent errors primarily affect the qubit population rather than the cavity photon number, leading to a higher error rate for the transmon. So the transmon is expected to experience greater error than the cavity.

For the even-parity initial state, the outcome remains unchanged with the change of the phase (\figpanels{fig:wig_13}{f}{h}), except for the rotation of all the states by $0.53\pm0.03$~rad that is found by maximizing the fidelity to the $(1/\sqrt{2})(\ket{2}+\ket{4})$ state. Part of the rotation, \unit[0.25]{rad}, can be attributed to cavity Kerr evolution during the two-photon drive. The remaining rotation likely results from other contributions to the Hamiltonian, such as Stark shifts induced by the drives and drive-dependent higher-order Kerr effects. The code words undergo a deterministic rotation under the action of the error-correcting drives, which can be compensated for by rotating the cavity states (or updating the reference frame of the cavity drive), conditioned on the outcome of the qubit measurement.

\subsection{Scaling SNAPPA }
It is important to be able to drive more transitions without significantly impacting the fidelity of the operation. This would allow our codewords to contain higher Fock states.  Importantly, the SNAPPA gates can be scaled by adding additional drives without a significant overhead in calibration or significant degradation of fidelity. In \cite{SuppMat}, we present a calibration procedure that adds small overhead when adding additional drives. We demonstrate the case of three simultaneous drives to the cavity where we have added a drive for the transition $\ket{5}\ket{g}\rightarrow \ket{6}\ket{e}$. We prepare $(\ket{1}+\ket{3}+\ket{5})\ket{g}/\sqrt{3}$ shown in \figpanel{fig:wig_135}{a}. Qubit population and fidelity after state preparation are given by 0.05 and 0.92, respectively, shown in \tabref{tab:Qubit_populations_and_fidelities}. Applying the off-resonant drives, the qubit population becomes 0.90, meaning that the SNAPPA adds 0.05 error to qubit population. We apply three cavity drives and one qubit drive. Now the phase of the second cavity drive at frequency $\omega_c+\Delta-4\chi$ directly maps onto the phase of Fock state $\ket{4}$ and the phase of the third cavity drive at frequency $\omega_c+\Delta-6\chi$ maps onto the phase of Fock state $\ket{6}$. Phases of the qubit drive and the first cavity drive at frequency $\omega_c+\Delta-2\chi$ are always kept at 0 as reference. We show four states in \figpanels{fig:wig_135}{b}{d}, namely $(1/\sqrt{3})(\ket{2}+\ket{4}e^{i\theta_{3\rightarrow4}}+\ket{6}e^{i\theta_{5\rightarrow6}})\ket{e}$, where $(\theta_{3\rightarrow4},\theta_{5\rightarrow6})$ is equal to $(0,0)$, $(\pi,0)$, $(0,\pi)$ and $(\pi,\pi)$  respectively. The state fidelity is equal for all four states (0.93) and the error introduced by the SNAPPA gate to the qubit state population is 0.05. This is similar to the errors introduced by the SNAPPA gate in~\figref{fig:wig_13} that is driving one less transition.  

\begin{figure}[h!]
\includegraphics[width=90mm]{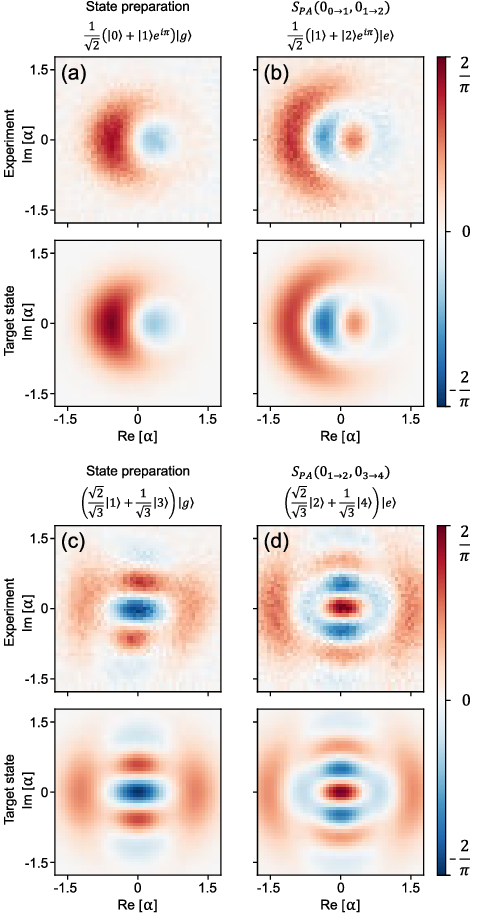}
\caption{\label{fig:wig_01} Wigner functions of experimental and target states. (a) Initial state $(\ket{0}+\ket{1}e^{i\pi})\ket{g}/\sqrt{2}$ and (b) $\ket{1}+\ket{2}e^{i\pi})\ket{e}/\sqrt{2}$, when drives $\ket{0}\ket{g}\rightarrow \ket{1}\ket{e}$ and $\ket{1}\ket{g}\rightarrow \ket{2}\ket{e}$ are applied. (c) Initial state $(\sqrt{2}/\sqrt{3}\ket{1}+1/\sqrt{3}\ket{3})\ket{g}$ and (d) state $(\sqrt{2}/\sqrt{3}\ket{2}+1/\sqrt{3}\ket{4})\ket{e}$, when drives $\ket{1}\ket{g}\rightarrow \ket{2}\ket{e}$ and $\ket{3}\ket{g}\rightarrow \ket{4}\ket{e}$ are applied.}
\end{figure}

In \figref{fig:wig_01} we demonstrate one more application of the SNAPPA gate. One can drive transitions for {\em all} photon number $n$ smaller than a maximum number to implement deterministic photon addition~\cite{Um2016}. This operation is a sibling to the probabilistic photon creation operator $a^\dagger$ often referred to as "single-photon addition" in quantum optics \cite{Parigi2007}. 
As an example, we apply the SNAPPA gate $S_{\rm PA}(0_{0\rightarrow1},0_{1\rightarrow2})$ by drives $\ket{0}\ket{g}\rightarrow\ket{1}\ket{e}$ and $\ket{1}\ket{g}\rightarrow\ket{2}\ket{e}$ on the initial state $(\ket{0}+\ket{1}e^{i\pi})\ket{g}/\sqrt{2}$ shown in \figpanel{fig:wig_01}{a} with qubit population 0.02 and cavity state fidelity 0.93 in \tabref{tab:Qubit_populations_and_fidelities}. Thus the cavity state becomes $(\ket{1}+\ket{2}e^{i\pi})/\sqrt{2}$ (\figpanel{fig:wig_01}{b}), with the qubit population 0.94 and the state fidelity 0.92 (\tabref{tab:Qubit_populations_and_fidelities}).
Another example is shown in \figpanels{fig:wig_01}{c}{d} for unequal superpositions of the two Fock states. Preparing the cavity in the state$\left(\sqrt{2}/\sqrt{3}\right)\ket{1} + \left(1/\sqrt{3}\right)\ket{3}$ (\figpanel{fig:wig_01}{c}), we apply the same drives as in the case of \figpanel{fig:wig_13}{b} to create $\left(\sqrt{2}/\sqrt{3}\right)\ket{2} + \left(1/\sqrt{3}\right)\ket{4}$(\figpanel{fig:wig_01}{d}) with qubit population 0.91 and state fidelity 0.96 (\tabref{tab:Qubit_populations_and_fidelities}).

\section{Discussion}
By using off-resonant drives, we have demonstrated the ability to selectively climb the Fock ladder and add arbitrary phases to the Fock states whose number is increased. Multiple such operations can be performed simultaneously. 
Similar to SNAP gates, SNAPPA gates need to be much longer than the inverse of the dispersive shift between the qubit and the cavity ($t_{\rm SNAPPA}\gg1/\chi$). However, it is likely possible to optimize the envelope shapes of off-resonant drives in order to shorten the gate time, as demonstrated for SNAP gates~\cite{PRXQuantum.3.030301}. Furthermore, using the third level of the qubit, SNAP gates were made path independent~\cite{reinhold2020error,Path_independant} which could possibly be adapted to make SNAPPA gates path independent as well.
It is also straightforward to implement a gate that selectively subtracts a photon while adding an arbitrary phase and exciting the qubit, as in the transformation $\ket{n+1}\ket{g}\rightarrow e^{i\theta_n}\ket{n}\ket{e}$. To drive this transition, we need drives whose frequencies satisfy $\omega_{qs}-\omega_{cs}=\omega_q-\omega_c-n\chi$. Keeping the qubit off-resonant drive the same, as in the case of SNAPPA gates ($\omega_q-\Delta$), we should further apply $n$ cavity off-resonant drives at frequencies $\omega_c-\Delta+n\chi$ to implement selective number-dependent arbitrary phase photon subtraction (SNAPPS). These drives can be combined with SNAPPA drives to give more functionality.


In our protocol, the SNAPPA drives are designed to selectively convert odd-parity states into even-parity states while leaving even-parity states unaffected. Regardless of the initial state’s parity, the combined effects of the Stark shift and Kerr nonlinearity result in a rotation of the Wigner function. For odd-parity states, we compensate for this rotation by optimizing the phases of the SNAPPA drives, which selectively targets the odd-parity states. In contrast, for even-parity states, no compensation is done, as the SNAPPA drives are intentionally configured to avoid interaction with these states, thereby maintaining the selective nature of our protocol. Importantly, in the context of error correction, we have the flexibility to choose the reference phase. This allows us to align the phase of the corrected state with the phase it would have acquired deterministically in the absence of errors. Consequently, the error correction process ensures that the final state retains the expected phase, effectively mitigating any unwanted phase deviations. However, this argument is valid when considering a superposition of only two Fock states, as in \figref{fig:wig_13}, where the Kerr effect simply introduces a global phase that can be factored out while preserving the relative phase between the states. For larger superpositions, such as in \figref{fig:wig_135}, the Kerr effect leads to different phases on each Fock component. In the case of odd-to-even parity transitions, SNAPPA corrects the phases on odd states caused by Kerr since we are actively addressing the odd-parity states. For even-parity states, however, the Kerr effect cannot be corrected by a simple rotation of the reference frame. Instead, it could be mitigated by applying a selective SNAP gate, as SNAP gates are specifically designed to apply phase corrections to individual Fock states.

To further improve SNAPPA, several aspects could be explored. First, improving the qubit lifetime would significantly enhance the fidelity of the gate~\cite{ni2022beating}. Second, optimizing the pulse shapes and parameters to minimize qubit-induced errors, such as relaxation, dephasing, and leakage, could enhance the fidelity of the gate. Third, developing techniques that decouple the qubit from the cavity more efficiently after the gate operation could mitigate transmon-induced decoherence in subsequent steps. Additionally, implementing an active or passive qubit reset immediately after the SNAPPA gate would prevent qubit population errors from propagating into the Wigner tomography. More generally, alternative state-reconstruction methods that do not rely on direct qubit involvement, such as using auxiliary qubits for indirect measurement~\cite{daSilva2009quantum, Gomes2016quantum}, could further reduce errors and are an important direction for future work.

The error recovery map probably could also be realized by using gradient-based optimization protocols such as GRAPE~\cite{eickbusch2022fast, heeres2017implementing, Puviani2025}. However, such black-box optimization will likely be very sensitive to system parameter calibration (including detailed stark shifts depending on precise driveline transmittance). In contrast, SNAPPA provides an interpretable calibration procedure.

In principle, the SNAPPA gate presented in \eqref{SNAPPA_gate} could be decomposed into a sequence of displacement and SNAP gates or via full optimal control such as GRAPE. However, in this work, we focus on the direct implementation of the exact recovery map in \eqref{SNAPPA_gate}, using the control drives and approach outlined in this paper and leave a detailed comparison for future work. We limit our comparison to noting, i) incorporating displacement gates into the unitary process likely leads to temporary population of higher Fock states, which might make the protocol more susceptible to noise, ii) GRAPE-based optimization might be very sensitive to parameter calibration in contrast to our implementation of SNAPPA where the gate is tuned up in a systematic iterative way.

In the context of quantum error correction, the SNAPPA drives can autonomously detect and correct changes in the parity of an initial superposition of Fock states, due to single-photon loss. To complete the error correction cycle, qubit reset must be performed, which can be implemented unconditionally \cite{magnard2018fast} or with measurement and feedback \cite{riste2012b}. More generally, SNAPPA gates make it possible to realize a family of non-unitary operations on the cavity, assisted by the ancillary qubit. These operations may be useful in the context of quantum algorithms to implement block encodings \cite{rall2020}.

\begin{acknowledgments}
We would like to thank Mats Myremark and Lars J\"onsson for machining the cavity. SG would like to thank Matteo Lostaglio for useful discussions. AME would like to thank Giulia Ferrini for useful discussions. The simulations and visualization of the quantum states were performed using QuTiP~\cite{Johansson2012,Johansson2013}, NumPy~\cite{Harris2020array}, and Matplotlib~\cite{Hunter2007}. The automatic differentiation tools TensorFlow~\cite{TensorFlow2015} and Jax~\cite{Jax2018} were used in state reconstruction and optimization. This work was supported by the Knut and Alice Wallenberg foundation via the Wallenberg Centre for Quantum Technology (WACQT) and by the Swedish Research Council. The chips were fabricated in the Chalmers Myfab cleanroom.
\end{acknowledgments}

\bibliography{apssamp}
\onecolumngrid
\clearpage

\input{SuppMat}

\end{document}

%% file: SuppMat.tex
\setcounter{section}{0}
\renewcommand{\thesection}{S\arabic{section}}
\setcounter{equation}{0}
\renewcommand{\theequation}{S\arabic{equation}}
\setcounter{figure}{0}
\renewcommand{\thefigure}{S\arabic{figure}}
\setcounter{table}{0}
\renewcommand{\thetable}{S\arabic{table}}
\renewcommand{\bibnumfmt}[1]{[S#1]}

\setcounter{page}{1}

%

\section{Hamiltonian parameters}
\label{app:A}

\begin{table}[h]
\caption{\label{tab:Hamilton_param}
Parameter values for the system Hamiltonian and measured coherence times. }
\begin{ruledtabular}
\begin{tabular}{ccc}
Parameter & Symbol & Value \\ 
\hline
Qubit frequency & $\omega_q/2\pi$ & \unit[5.523]{GHz} \\
Cavity frequency & $\omega_c/2\pi$ & \unit[3.581]{GHz} \\ 
Resonator frequency & $\omega_r/2\pi$ & \unit[7.815]{GHz} \\
Qubit-cavity disp.~shift & $\chi_{qc}/2\pi$& \unit[1.44]{MHz}\\
Qubit-resonator disp.~shift & $\chi_{qr}/2\pi$& \unit[0.8]{MHz} \\
Cavity Kerr coeff. & $K_c/2\pi$& \unit[2.2]{kHz}  \\
Qubit anharmonicity & $\alpha_{q}/2\pi$& \unit[231]{MHz} \\
Qubit-cavity disp.~shift corr. & $\chi'_{qc}/2\pi$& \unit[3]{kHz}\\
Qubit relaxation time & $T_{1q}$ & \unit[80 $\pm$ 6]{$\mu$s} \\
Qubit decoherence time & $T_{2q}$ & \unit[20 $\pm$ 3]{$\mu$s} \\
Cavity relaxation time & $T_{1c}$ & \unit[567 $\pm$ 13]{$\mu$s} \\
\end{tabular}
\end{ruledtabular}
\end{table}

The system parameters are listed in \tabref{tab:Hamilton_param}. The different parameters such as the dispersive shift $\chi$, the correction to the dispersive shift $\chi'$, and the Kerr nonlinearity of the cavity are measured following the calibration experiments in Ref.~\cite{Reinhold_PhD}. When doing state preparation we use two optimized SNAP gates each \unit[1]{$\mu$s} long, interleaved by three displacements each \unit[50]{ns} long~\cite{PRXQuantum.3.030301}. The sampling frequency of an intermediate frequency signal was \unit[1]{GSa/s}. This signal gets digitally up-converted to appropriate GHz frequency, passes digital to analog converter and is sent to the sample.

\section{Calibration of the off-resonant drive amplitudes and phases}
\label{app:calibration_amp_phase}

\begin{figure*}[h!]
\includegraphics[width=180mm]{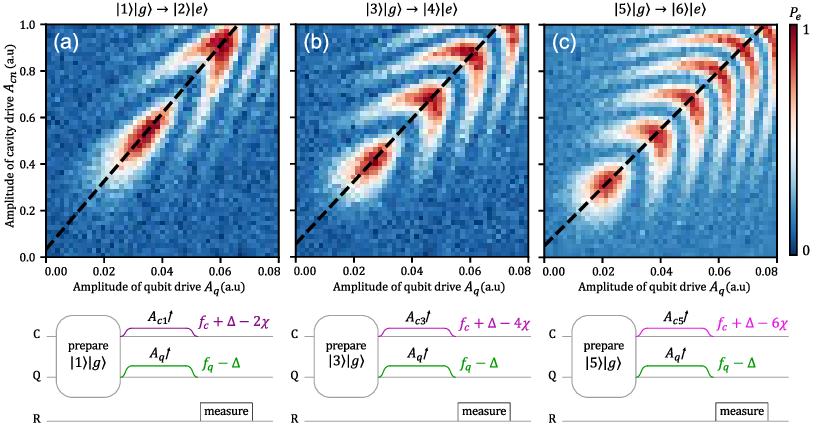}
\caption{\label{fig:amp_cal}Amplitude Rabi-like measurement for calibrating the approximate amplitudes of the off-resonant drives. The 
frequency of the qubit drive is $f_q-\Delta$ ($\Delta$=\unit[30]{MHz}) in all of the panels. The frequency of the cavity drive is (a) $f_c+\Delta-2\chi$, (b) $f_c+\Delta-4\chi$, and (c) $f_c+\Delta-6\chi$. The black dashed lines on top of the data 
are given by the following relations: (a) $A_c=14.62A_q+0.03$, (b) $A_c=13.46A_q+0.06$, and (c) $A_c=12.53A_q+0.05$. The pulse sequence corresponding to each panel is presented directly under the measurement result.}
\end{figure*}

\begin{figure*}[h!]
\includegraphics[width=180mm]{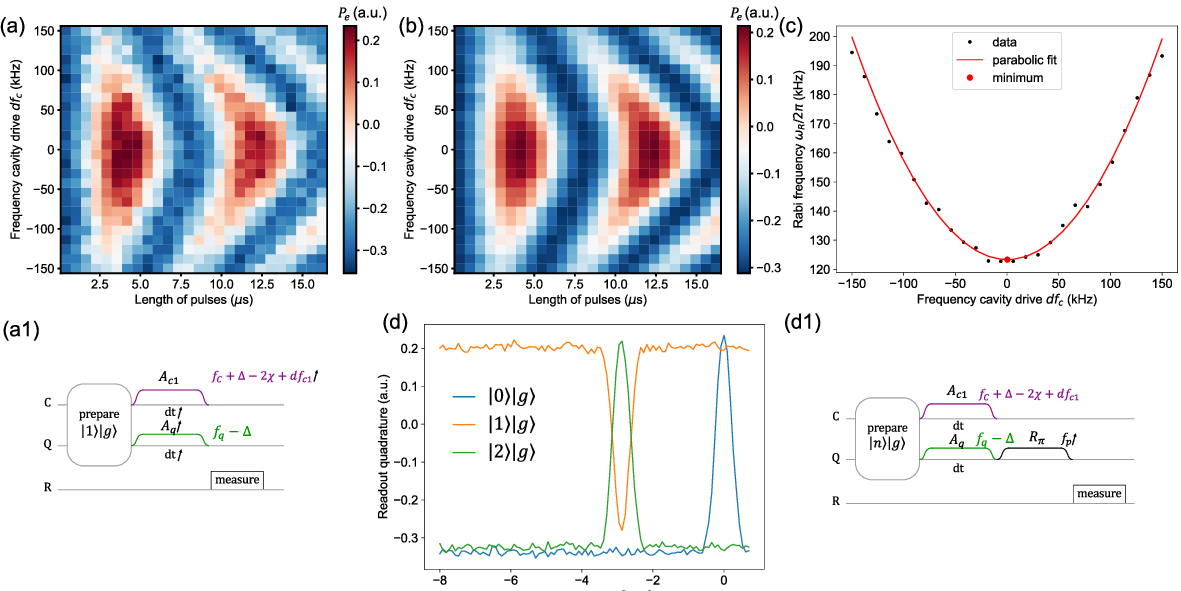}
\caption{\label{fig:calibrating_single_transition} Finding $\pi$ pulse for transition $\ket{1}\ket{g}\rightarrow\ket{2}\ket{e}$. (a) Rabi Chevron data for pulse sequence from (a1). (b) Fit and (c) parabolic fit to Rabi rate extracted from (b). (d) Check of the pulse we found in (a). We prepare different Fock states, apply the drives and check the population of the cavity (d1). As expected, preparing Fock states $\ket{0}$ or $\ket{2}$ nothing changes. When we prepare Fock state $\ket{1}$ the state of the system after applied drives is $\ket{2}\ket{e}$.}
\end{figure*}

\begin{figure*}[h!]
\includegraphics[width=180mm]{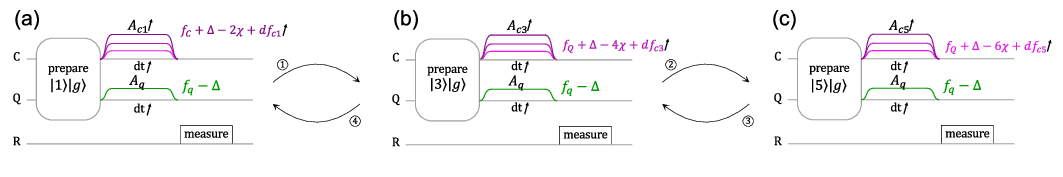}
\caption{\label{fig:pulse_sequence}Sequence of experiments to follow in order to calibrate amplitudes and frequencies of multiple cavity drives. We stop when the results converge. Starting from a good guess only six experiments are necessary to converge. }
\end{figure*}

\begin{figure*}[h!]
\includegraphics[width=140mm]{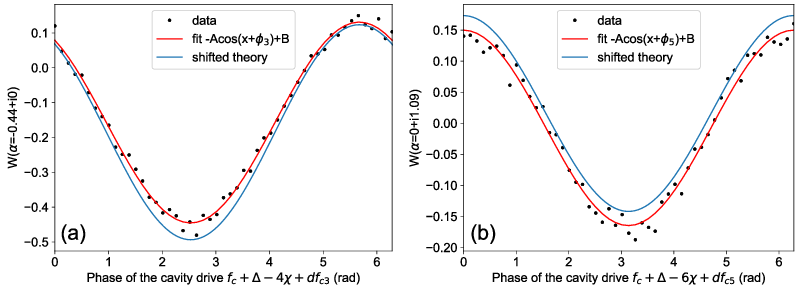}
\caption{\label{fig:phase_cal}Example of calibrating phases of the cavity drives when the qubit drive at frequency $f_q-\Delta$ and three cavity drives at frequencies $f_c+\Delta-(n+1)\chi, n\in\{1,2,3\}$ are simultaneously applied. We prepare the state (a) $1/\sqrt{2}(\ket{1}+\ket{3})\ket{g}$ ((b) $1/\sqrt{2}(\ket{1}+\ket{5})\ket{g}$) and apply all four drives while varying the phase of the drive at frequency (a) $f_c+\Delta-4\chi+df_{c3}$ ((b) $f_c+\Delta-6\chi+df_{c5}$). We measure the value of the Wigner function at the point (a) $\alpha=0.44+i0$ ((b) $\alpha=0+i1.09$). We fit a cosine function (in red) where (a) $\phi_4=-0.61$ ((b) $\phi_6=-3.15$) tells us the shift from our chosen phase reference point where Fock state $\ket{1}$ has phase 0. The theoretical value of Wigner function of the state (a) $1/\sqrt{2}(\ket{2}+\ket{4}e^{i(\phi-\phi_3)})$ ((b) $1/\sqrt{2}(\ket{2}+\ket{6}e^{i(\phi-\phi_5)})$) in the point (a) $\alpha=0.44+i0$ ((b) $\alpha=0+i1.09$) is shown in blue. }
\end{figure*}

The procedure we follow to calibrate the amplitudes and the phases of the off-resonant drives has several steps and is general no matter how many drives we want to calibrate. We will give a specific example corresponding to the calibration of the drives
used in Fig.~3 of the main text. 
For this particular case, we want to apply the
SNAPPA gate $S_{\rm PA}(0_{1\rightarrow2},\zeta_{3\rightarrow4},\xi_{5\rightarrow6})$. This means we have four pulses to apply. The qubit off-resonant pulse at frequency $f_{qs}=f_q-\Delta$ with amplitude $A_q$ and three cavity off-resonant pulses at frequencies $f_{csn}=f_c+\Delta-(n+1)\chi+df_{cn},n\in\{1,3,5\}$ with amplitudes $A_{cn},n\in\{1,3,5\}$. We pick the length of the drives to be \unit[4.2]{$\mu$s}. We have four amplitudes $A_q$, $A_{c1}$, $A_{c3}$, and $A_{c5}$, and three frequencies $df_{c1}$, $df_{c3}$, and $df_{c5}$. We choose to keep the phases of the qubit off-resonant drive at frequency $f_{qs}$ and the cavity off-resonant drive at frequency $f_{cs1}$ constant. Thus we have two phases to calibrate. The phase of the drive $f_{cs3}$ ($\zeta$) and $f_{cs5}$ ($\xi$). Here is the procedure:\newline

1. Perform an amplitude Rabi-like measurement (\figref{fig:amp_cal}) for all the transitions we want to drive. We apply a qubit pulse at frequency $f_{qs}=f_q-\Delta$ ($\Delta=$~\unit[30]{MHz}) and a cavity drive at frequency $f_{csn}=f_c+\Delta-(n+1)\chi$, with $n=1$, $n=3$, and $n=5$ in \figref{fig:amp_cal} panel (a), (b), and (c) respectively. Find the linear relation between $A_{cn}=aA_q+b$. This gives us an approximate idea of the amplitudes of the $\pi$ pulses. \newline

2. Find the $\pi$ pulse for the lowest Fock state we want to drive. We do this by sweeping the length of the drives and the frequency of the cavity drive for given amplitude $A_q$ and $A_c=aA_q+b$ and measure the state of the qubit (\figpanel{fig:calibrating_single_transition}{a}). We find the length of the $\pi$ pulse and the $df_c$(\figpanel{fig:calibrating_single_transition}{c}), where $df_c$ is the Stark shift correction of the cavity drive $f_sc=f_c+\Delta-n\chi+df_c$.\newline

3. Similar like previous point, but now we keep all the desired drives on (\figref{fig:pulse_sequence}). Starting from the parameters found in 2. and a good guess for the remaining two drives (from \figpanel{fig:amp_cal}{b} and (c)) we follow the experiments depicted in (\figref{fig:pulse_sequence}). We prepare Fock state $\ket{1}$ and find the amplitude $A_{c1}$ and frequency shift $df_{c1}$ (\figpanel{fig:pulse_sequence}{a}). Then with the knowledge of those parameters updated we prepare Fock state $\ket{3}$ and find the amplitude $A_{c3}$ and frequency shift $df_{c3}$ (\figpanel{fig:pulse_sequence}{b}). Next, we repeat the process for Fock state $\ket{5}$ and back to Fock state $\ket{3}$ and $\ket{1}$ and so on until the parameters we are extracting converge. With a good initial guess it took 6 steps until the parameters converged. \newline 

4. Finally we calibrate the phases of drives $f_{cs3}=f_c+\Delta-4\chi+df_{c3}$ and $f_{cs5}=f_c+\Delta-6\chi+df_{c5}$, $\zeta$ and $\xi$ respectively (\figref{fig:phase_cal}). We prepare the state $(1/\sqrt{2})(\ket{1}+\ket{3})\ket{g}$ and apply all the drives. We sweep the phase of the drive at frequency $f_{cs3}=f_c+\Delta-4\chi+df_{c3}$ and measure the value of Wigner function in a point $\alpha=0.44+i0$. We pick this point because The Wigner function of created state $(1/\sqrt{2})(\ket{2}+\ket{4}e^{i\zeta})\ket{e}$ has good contrast (\figpanel{fig:phase_cal}{a}). We fit a cosine function in order to find the phase shift $\phi_3$ between the drive phase and the reference frame where the phase of Fock state $\ket{1}$ is zero. Similarly, we find the phase of drive $f_{cs5}=f_c+\Delta-6\chi+df_{c5}$ $\phi_5$ by preparing the state $(1/\sqrt{2})(\ket{1}+\ket{5})\ket{g}$, applying the drives and measuring Wigner point $\alpha=0+i1.09$ (\figpanel{fig:phase_cal}{b}).\newline

\section{Effective Hamiltonian Modeling}

Here we give a general overview of the derivation of the effective Hamiltonian for the SNAPPA gate, starting from our system Hamiltonian
\be \label{H_two_drives}
H(t) = \omega_q q^\dag q + \omega_c a^\dag a - \frac{E_J}{4!} \phi^4 + \epsilon_1 \cos(\omega_1 t) \left( q + q^\dag \right)
+ \epsilon_2 \cos(\omega_2 t) \left( a + a^\dag \right),
\ee
where $\omega_c$ and $\omega_q$ are the frequency of cavity and qubit, and $\phi = \varphi_q (q + q^\dag) + \varphi_c (a + a^\dag)$ with $\varphi_q$ and $\varphi_c$ the participation ratios of the qubit and cavity. Here $\epsilon_1$ and $\epsilon_2$ are the strengths of the qubit and cavity drives, respectively, with frequencies $\omega_1$ and $\omega_2$.

To capture the two-photon interaction driven by the SNAPPA gate, we need the \( b^\dagger a^\dagger \)-interaction in the effective Hamiltonain as described in~\cite{Campagne-Ibarcq2018}. However, to ensure that the effective Hamiltonian aligns with experimentally measured data, additional terms must be included, as outlined in~\cite{Jirlow2025}. 

Our modeling approach begins by expanding the transmon cosine potential to fourth order and retaining all terms from this expansion. We then move into a frame rotating at the drive frequencies, followed by a displacement transformation. The displacement amplitudes are chosen such that they cancel the drive terms in the Hamiltonian. Finally, we apply a rotating wave approximation (RWA) to arrive at the effective model used in this work.

\begin{figure}
\centering
\includegraphics[width=0.8\columnwidth]{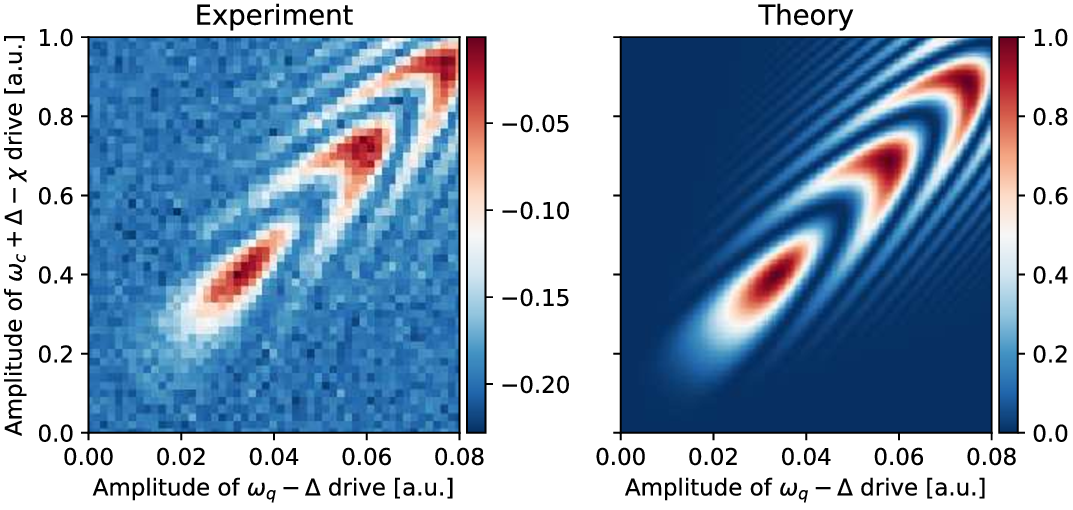}
\caption{Qubit population after driving the $\ket{0}\ket{g}\rightarrow\ket{1}\ket{e}$ transition with varying drive amplitudes, and a gate time of $\tau = 4.2\ \mu s$.  Experimental data shows excellent agreement with the theoretical model.}
\label{Exp_vs_Th}
\end{figure}

As shown in~\figref{Exp_vs_Th}, the theoretical model shows excellent agreement with experimental results when SNAPPA drives are applied to the $\ket{0}\ket{g} \rightarrow \ket{1}\ket{e}$ transition across varying drive amplitudes.

We aim to model the $b^\dagger a^\dagger$-interaction as in \cite{Campagne-Ibarcq2018}, which describes the two-photon interaction driven by the gate. However, for the effective Hamiltonian to agree with experimentally measured data we need to include additional terms, as described by \cite{Jirlow2025}. We model our system by expanding the transmon cosine potential to the fourth order, and then expanding the fourth order term, keeping all terms. We then move into a frame rotating at the drive frequencies, followed by a displacement transformation. The displacement amplitudes are chosen to cancel the drive terms in our drive Hamiltonian. Finally we apply a rotating wave approximation, to arrive at our effective model.

\section{Wigner tomography analysis in the presence of residual population in qubit}

We analyze the Wigner tomography procedure for a cavity coupled to a transmon ancilla. This method requires the transmon to be initialized in its ground state. The tomography process begins with a $\hat{Y}_{\pi/2}$ rotation on the qubit, followed by free evolution under the dispersive shift, implementing a conditional phase gate $\hat{C}_\Phi$ of the form
\begin{equation} \hat{C}_\Phi = \mathcal{I} \otimes \ketbra{g}{g} + e^{i\Phi \hat{a}^\dag \hat{a}} \otimes \ketbra{e}{e}, \end{equation}
where $\Phi = t\chi$. By setting $t = \frac{1}{2\chi}$, odd Fock states acquire a $\pi$ phase shift, while even Fock states remain unchanged. A second $\hat{Y}_{\pi/2}$ rotation maps this parity information onto the ground and excited states of the transmon, which can be measured via standard qubit readout. Applying this sequence to an initial cavity state $\ket{\psi}$, the overall transformation $\hat{W}$ is
\begin{equation}
    \hat{W}\ket{\psi, g} = \hat{Y}_{\pi/2}\hat{C}_\pi \hat{Y}_{\pi/2} \ket{\psi, g} = \hat{\Pi}_{\text{odd}}\ket{\psi, e} + \hat{\Pi}_{\text{even}}\ket{\psi, g},
\end{equation}
where $\hat{\Pi}_{\text{odd}}$ and $\hat{\Pi}_{\text{even}}$ project onto the odd and even Fock state subspaces, respectively. This entangles the qubit with the photon-number parity of the cavity, such that measuring the transmon in $\ket{e}$ corresponds to an odd photon-number parity, and vice versa. Repeating this process over many trials allows us to estimate the expectation value of the photon-number parity at different displacements $\alpha$, thereby reconstructing the Wigner function
\begin{equation} W(\alpha) = \frac{2}{\pi} \mathrm{Tr} \mleft[ { \hat{D}^\dag_\alpha \rho \hat{D}_\alpha \hat{P} } \mright], \end{equation}
where $\hat{P}$ is the photon-number parity operator.

We now analyze errors in Wigner tomography arising from initial population in the transmon. Suppose the transmon and the cavity are initially in an entangled state
\begin{equation}
    \ket{\phi_I} = \alpha \ket{\psi, g} + \beta\ket{\phi, e},
\end{equation}
where $\ket{\psi}$ and  $\ket{\phi}$ are arbitrary cavity states. Applying $\hat{W}$ to this state yields
\begin{equation}
    \hat{W}\ket{\phi_I} = \alpha\Big(\hat{\Pi}_{\text{odd}}\ket{\psi, e} + \hat{\Pi}_{\text{even}}\ket{\psi, g}\Big) + \beta\Big(\hat{\Pi}_{\text{even}}\ket{\phi, e} + \hat{\Pi}_{\text{odd}}\ket{\phi, g}\Big).
\end{equation}
The tomography error probability $p_{\text{err}}$ corresponds to measuring $\ket{e}$ when the cavity is in an even-parity state, or $\ket{g}$ when it is in an odd-parity state. This requires that the qubit excited state is entangled with the even parity projector, and the ground state is entangled with the odd parity projector. It should be noted that this definition of tomography error just includes the probability of measuring the incorrect parity of the cavity. It does not make any assumptions about the cavity state, or assume that there is some specific cavity state that is the only correct state to measure. The projection operator onto this error subspace is
\begin{equation}
    \hat{\Pi}_{\text{err}} = \hat{\Pi}_{\text{even}} \otimes \ket{e}\bra{e} + \hat{\Pi}_{\text{odd}} \otimes \ket{g}\bra{g}.
\end{equation}
Thus, the probability of an incorrect parity measurement, irrespective of the cavity state, is
\begin{equation}
    p_{\text{err}} = \bra{\phi_I}\hat{W}^\dag \hat{\Pi}_{\text{err}} \hat{W}\ket{\phi_I} = |\beta|^2,
\end{equation}
which equals the initial population of the transmon’s excited state. Similarly, for an initial mixed state
\begin{equation}
    \rho_I = (1 - |\beta|^2)\ket{\psi, g}\bra{\psi, g} + |\beta|^2\ket{\phi, e}\bra{\phi, e}
\end{equation}
the error probability is simply $p_{\text{err}} = |\beta|^2$. This analysis shows that any initial population in the transmon directly translates into an error in Wigner tomography, emphasizing the importance of initializing the transmon in its ground state before performing parity measurements.